\documentclass[preprint,showpacs,showkeys,groupedaddress,superscriptaddress]{revtex4}
\usepackage{epsf,epsfig}
\usepackage[psamsfonts]{amssymb}
\usepackage{amsmath}
\usepackage{bm}
\usepackage{natbib}

\begin{document}

\title{A perturbative approach to calculating the Casimir force in fluctuating scalar and vector fields}

\author{F. Kheirandish  }
\email{fardin$_$kh@phys.ui.ac.ir}
\author{ M. Jafari }
\email{jafary.marjan@gmail.com} \affiliation{Department of Physics, Faculty of Science, University of Isfahan,
Hezar-Jarib St., 81746-73441, Isfahan, Iran}

\date{\today}

\begin{abstract}
\noindent Based on a perturbative approach, a series expansion in
susceptibility function of the medium is obtained for the Casimir
force between arbitrary shaped objects foliated in a scalar or
vector fluctuating field in arbitrary dimensions.
Finite-temperature corrections are derived and the results are
compared in first order with weak coupling regime in scattering
method. The generalization to a massive vector field is also
investigated.
\end{abstract}


\maketitle

\section{Introduction}\label{Introduction}
\noindent The path integral is a formulation of quantum mechanics equivalent to the standard formulations, offering a new way of looking at the
subject which is more intuitive than the usual approaches.
Application of the path integrals are as vast as quantum mechanics
itself, including the quantum mechanics of a single particle,
statistical mechanics, condensed matter physics and quantum field
theory \cite{Greiner,kleinert}. Quantum field theory is the
quantum mechanics of continuous systems and fully developed in
quantum electrodynamics which is the most successful theory in
physics. The path integral in quantum field theory, both
relativistic and non-relativistic, play a much more important
role, for several reasons. They provide a relativity easy road to
quantization and to expressions for Green's functions, which are
closely related with many physical quantities. The close relation
between statistical mechanics and quantum field theory is plainly
visible via path integrals. Usually we are interested in a quantum
field which has to be considered in the presence of a matter field
described by some bosonic fields. For example, in quantum optics
there are some problems where we need to quantize the electromagnetic field in the presence of some dielectrics
or in calculating the effects of matter fields on Casimir forces
\cite{3,5,matloob,kheirandish}. For a consistent quantization of such fields we should include the matter fields in the process of quantization.
But the main problem is that there are few situations where the interested
physical quantities, like for example, the Casimir force, can be
determined in a closed form and finding an effective approximation
method is quite necessary \cite{Lifshitz}.

The Casimir energy is the difference between the energy of the fluctuating field when then objects are present
and when the objects are removed to infinity. The advent of precision experimental measurements of Casimir forces
 \cite{emig1,emig2,emig3,emig4,emig5} and the possibility that they can be applied to nanoscale electromechanical devices
  \cite{emig6,emig7} has stimulated interest in developing a practical way to calculate
the dependence of Casimir energies on the shapes of the objects. Many geometries have been analyzed over the years,
but the case of compact objects has proved rather difficult.
 One way to calculate the Casimir effect is using the multiple scattering formalism, which dates back at least to the 1950.
Recently Milton and et al., have noticed that the multiple
scattering method can yield exact, closed form results for bodies
that are weakly coupled to the quantum field
\cite{milton1,milton2}.

In the present work, we use path integrals to calculate the Casimir
force between arbitrary shaped objects foliated in a scalar or
vector fluctuating field. For this purpose, we start from a
Lagrangian and develop a perturbative approach to calculating the
Casimir force in the case of scalar and vector fields by taking
into account the susceptibility function of the medium as the
expansion parameter. The covariant formulation of the method and
also its finite-temperature correction is given which may have
applications in dynamical Casimir effects
\cite{Amooshahi,Golestanian}. Finally, it is shown that the
results are consistent with the weak coupling limit in multiple
scattering formalism as expected
\cite{milton1,milton2,milton3,milton4,milton5,milton6,milton7,milton8}. The present work is in fact a generalization of the previous work \cite{kheirandish1} as follows: (i) The massless scalar field in $1+1$-space time is fully investigated and an expression for Casimir force between intervals is derived. (ii) The Proca electromagnetic field is investigated and an expression for the free energy in zero and finite temperature is obtained for two arbitrarily shaped dielectrics. (iii) the approximations are improved and the results are compared with the weak coupling regime in the scattering approach.  

\section{Massless Scalar Field}\label{SCALAR FIELD}
\noindent The Lagrangian of a massless Klein-Gordon field in
$N+1$-dimensional space-time $(x=(\textbf{x},t)\in
\mathbb{R}^{N+1})$ is the simplest field theory given by
\begin{equation}\label{system lagrangian}
{\cal L}_s= \frac{1}{2}\partial_\mu \varphi(x) \partial^ \mu
\varphi(x).
\end{equation}
As usual, let the medium be modeled by a continuum of harmonic
oscillators, namely the Hopfield model \cite{hopfield} as follows
\begin{equation}\label{reservior lagrangian}
{\cal L}_{res}= \frac{1}{2}\int\limits _0 ^\infty d\omega \left(\dot{Y}_{\omega} ^2 (x)-
{\omega}^2 Y_{\omega}^2 (x) \right).
\end{equation}
This modeling can be easily generalized to a covariant form but
let us first discuss the non relativistic case. The interaction
between the scaler field and its medium is assumed to be linear
and described by
\begin{equation}\label{interaction lagrangian}
{\cal L}_{int}=\int \limits_0 ^\infty d\omega f(\omega,\textbf{x})\dot{Y}_\omega (x)\varphi(x),
\end{equation}
where $f(\omega,\textbf{x})$ is the coupling function between the
scaler field and its medium. If the medium has filled a finite
space $\Omega$, then for $(\textbf{x}\notin \Omega)$, there is no
matter to be coupled to the field so we should set
$f(\omega,\textbf{x})=0$. Having the total Lagrangian, we can
quantize the total system using path-integral techniques
\cite{greiner1}. An important quantity in any field theory is the
generating functional from which n-point correlation functions can
be obtained by taking successive functional derivatives. Here our
purpose is to find two-point correlation functions or Green's
function in terms of the susceptibility of the medium. For this
purpose the free generating-functional can be written as
\cite{kheirandish1}
\begin{eqnarray}\label{final free generation functional}
W_0[J,J_\omega]&=&Ne^{-\frac{1}{2 \hbar^2} \int d^n\textbf{x} \int dt \int d^n\textbf{x}'
\int dt' J(x)G^0(x-x')J(x')} \nonumber \\
&\times& e^{-\frac{1}{2 \hbar^2} \int d^n\textbf{x} \int dt \int d^n\textbf{x}' \int dt'
\int \limits _0 ^\infty d\omega J_\omega(x)G_\omega^0(x-x')J_\omega(x')}.
\end{eqnarray}
From the free generating-functional we can obtain the interacting generating-functional from the well known formula 
\cite{greiner1}
\begin{eqnarray}\label{generating function}
W[J,J_\omega]&=&e^{\frac{i}{\hbar} \int d^n\textbf{x} \int dt \int \limits_0 ^\infty d\omega f(\omega,\textbf{x})
 (\frac{\hbar}{i} \frac{\delta}{\delta J(x)}) \frac{\partial}{\partial t} (\frac{\hbar}{i} \frac{\delta}
 {\delta J_\omega (x)})} W_0[J,J_\omega] \nonumber \\
&=&N e^{-i\hbar \int d^n\textbf{x} \int dt \int \limits_0 ^\infty d\omega f(\omega,\textbf{x}) \frac{\delta}{\delta J(x)}
\frac{\partial}{\partial t} \frac{\delta}{\delta J_\omega (x)}} \nonumber \\
 &\times & e^{-\frac{1}{2 \hbar^2} \int d^n\textbf{x} \int dt \int d^n\textbf{x}' \int dt' J(x)G^0(x-x')J(x')} \nonumber \\
 &\times& e^{-\frac{1}{2 \hbar^2} \int d^n\textbf{x} \int dt \int d^n\textbf{x}' \int dt' \int \limits _0 ^\infty
 d\omega J_\omega(x)G_\omega^0(x-x')J_\omega(x')}.
\end{eqnarray}
where
\begin{eqnarray}\label{Green functions}
G^0 (x-x') &= & i\hbar \int {\frac{d^n\textbf{k} d k_0}{(2 \pi)^{n+1}}} {\frac {e^{-i\textsl{k}(\textbf{x}-\textbf{x}')}
e^{ik_0(t-t')}}{k_0^2-\textbf{k}^2+i\epsilon}},\hspace{1cm}(\epsilon>0), \nonumber \\
G_\omega ^0 (x-x') &=& i\hbar \delta^n(\textbf{x}-\textbf{x}')
\int {\frac{d k_0}{2 \pi}} {\frac {e^{ik_0(t-t')}}{(k_0)^2- \omega
^2 +i\epsilon}},
\end{eqnarray}
Now having the generating functional, the two-point function or
Green's function can be obtained as
\begin{equation}\label{2 point Green function}
G(x-x')=(\frac{\hbar}{i})^2 \frac{\delta ^2}{\delta J(x) \delta J(x')} W[J,J_\omega]\mid_{j,j_\omega=0}.
\end{equation}
By using Eq.(\ref{generating function}) and after some
straightforward calculations, we find the following expansion for
Green's function in frequency variable
\begin{eqnarray}\label{expansion Green function}
G(\textbf{x}-\textbf{x}',\omega)=G^0(\textbf{x}-\textbf{x}',\omega)+ \int_\Omega d^n \textbf{z}_1 G^0(x-z_1,\omega)
[\omega^2 \tilde{\chi}(\omega,\textbf{z}_1)]G^0(\textbf{z}_1-\textbf{x}',\omega)+ \nonumber\\
\int_\Omega \int_\Omega d^n \textbf{z}_1 d^n \textbf{z}_2
G^0(\textbf{x}-\textbf{z}_1,\omega) [\omega^2
\tilde{\chi}(\omega,\textbf{z}_1)]G^0(\textbf{z}_1-\textbf{z}_2,\omega)
[\omega^2
\tilde{\chi}(\omega,\textbf{z}_2)]G^0(\textbf{z}_2-\textbf{x}',\omega)
+\cdots,
\end{eqnarray}
where $\tilde{\chi}(\omega,\textbf{x})$ is the susceptibility
function of the medium in frequency variable.

By using Euler lagrange equations we find equation of motion for
the fields and in particular we find a Langevin type equation for
the scalar field. Specially we can show that the Green's function
of the system satisfies
\begin{equation}\label{Green function}
\partial_\mu \partial^\mu G(\textbf{x}-\textbf{x}',t-t')+{\frac{\partial}{\partial t}} \int
\limits _{-\infty}^t dt'' \chi(t-t'',\textbf{x}) \frac{\partial}{\partial t''}
G(\textbf{x}-\textbf{x}',t''-t')=\delta(\textbf{x}-\textbf{x}',t-t').
\end{equation}
 We can also show that the coupling function between the scalar field and its medium is proportional to the imaginary
 part of the susceptibility which is responsible for dissipation in the system
\begin{equation}\label{coupling function}
f^2(\omega,\textbf{x})=\frac{\omega}{\pi}\Im[\chi(\omega,\textbf{x})].
\end{equation}
The memory function in a homogeneous medium is position independent and therefore, Eq.(\ref{Green function}) can be solved easily
in reciprocal space
\begin{equation}\label{reciprocal}
\tilde{G}(\textbf{k},\omega)=\frac{1}{\textbf{k}^2-\omega ^2-\omega ^2 \tilde{\chi}(\omega)}=
\frac{1}{\textbf{k}^2-\omega ^2 \frac{\epsilon(\omega)}{\epsilon_0}},
\end{equation}
where $\frac{\epsilon({\omega})}{\epsilon_0}=1+\tilde{\chi}(\omega)$ can be considered as the dielectric function of the
medium. From Eq.(\ref{reciprocal}) it is clear that Green's function in the presence of a homogeneous medium can be obtained
from Green's function of the free space simply by substituting $\omega^2$ with
$\omega^2 \frac{\epsilon(\omega)}{\epsilon_0}$. Now Eq.(\ref{Green function}) in frequency-space can be written as
\begin{equation}\label{Green function 2}
[\nabla^2 -\omega^2 \frac{\epsilon(\omega,\textbf{x})}{\epsilon_0}]G(\textbf{x}-\textbf{x}',\omega)=
\delta(\textbf{x}-\textbf{x}').
\end{equation}
In some simple geometries where dielectric function is a piecewise homogeneous function, for example for an array of
parallel dielectric slabs, Eq.(\ref{Green function}) has an exact solution, see for example \cite{matloob}.
The function $\epsilon(\omega,\textbf{x})$ in a general piecewise homogeneous medium is defined by
\begin{eqnarray}\label{epsilon}
\frac{\epsilon(\omega,\textbf{x})}{\epsilon_0}=\left\{
                                        \begin{array}{ll}
                                          {\epsilon(\omega)}, & \hbox{$\textbf{x}\in \Omega$} \\
                                          1, & \hbox{$\textbf{x}\notin \Omega$} \\
                                        \end{array}
                                      \right.
\end{eqnarray}
where $\Omega$ is the union of regions filled with a matter with the dielectric function $\epsilon({\omega})$. In general, finding an exact solution of Eq.(\ref{Green function 2}) is quite complicated or even impossible and a series solution will be useful. 
\section{Covariant formulation }\label{covarient formulation}
\noindent For obtaining a covariant formulation we model the medium by a continuum of Klein-Gordon fields and write the total 
Lagrangian density as follows
\begin{equation}\label{lagrangian total}
{\cal L}= {\cal L}_s +{\cal L}_{res}+{\cal L}_{int},
\end{equation}
where
\begin{eqnarray}\label{lagrangian}
{\cal L}_m&=& \frac{1}{2} \partial_\mu \varphi(x)\partial^\mu \varphi(x), \nonumber\\
{\cal L}_{res}&=& \frac{1}{2} \int \limits _0 ^\infty d\omega (\partial_\mu Y_\omega(x) \partial^\mu
Y_\omega (x)-\omega^2 Y^2_\omega (x)), \nonumber\\
{\cal L}_{int}&=& \int \limits _0^\infty d\omega f^{\mu}(\omega,x)Y_\omega(x) \partial_\mu \varphi(x),
\end{eqnarray}
are the Lagrangian densities of the scalar field, its medium and their
interaction, respectively. Here $x=(x^0,\textbf{x})$ belongs to
the Minkowski space-time $\mathbb{R}^{N+1}$. The fields
$\varphi(x)$ and $Y_\omega (x)$ are scaler fields and the coupling
function $f^\mu (\omega,x)$ is a vector field. From Euler-Lagrange
equations we find the Langevin equation for $\varphi(x)$
\begin{equation}\label{difrential equation covariant}
\partial_\mu \partial^\mu \varphi (x)+\partial_\mu \int d^{n+1}x' \chi^{\mu\nu}(x,x')\partial'_\nu \varphi(x')=J^N(x),
\end{equation}
and the Green's function of the system and the medium satisfy  the
following equations
\begin{eqnarray}\label{cov,difrential equation}
\frac{i}{\hbar}(\partial_\mu \partial^\mu)G^0 (x-x')=\delta(x-x') , \nonumber \\
\frac{i}{\hbar}(\partial_\mu \partial^\mu +\omega^2)G^0_\omega
(x-x')=\delta(x-x').
\end{eqnarray}
In Eq.(\ref{difrential equation covariant}) the noise source and
relativistic susceptibility are respectively defined by
\begin{equation}\label{source}
J_N(x)=-\partial_\mu \int \limits _0^\infty d\omega f^\mu(\omega,x)Y^N_\omega(x),
\end{equation}
\begin{equation}\label{relativistic susceptibility}
\chi^{\mu\nu}(x-x')=\int \limits_0^\infty d\omega f^\mu(\omega,x)G_\omega(x-x')f^\nu(\omega,x').
\end{equation}
Now using Eq.(\ref{lagrangian}) we can find two-point correlation
function or Green's function in terms of the susceptibility of
the medium. We have
\begin{eqnarray}\label{cov.generating function}
W[\rho,\rho_\omega]&=&e^{\frac{i}{\hbar} \int d^{n+1}x \int \limits_0 ^\infty d\omega f(\omega,x)
(\frac{\delta}{\delta \rho(x)}) {\partial_\mu} (\frac{\delta}{\delta \rho_\omega (x)})} W_0[\rho,\rho_\omega] \nonumber \\
&=&N e^{-\frac{i}{\hbar} \int d^{n+1}x \int \limits_0 ^\infty d\omega f(\omega,x) \frac{\delta}{\delta \rho(x)}
{\partial_\mu} \frac{\delta}{\delta \rho_\omega (x)}} \nonumber \\
 &\times & e^{-\frac{1}{2 \hbar^2} \int d^{n+1}x \int d^{n+1}x'  \rho(x)G^0(x-x')\rho(x')} \nonumber \\
 &\times& e^{-\frac{1}{2 \hbar^2} \int d^{n+1}x  \int d^{n+1}x' \int \limits _0 ^\infty
 d\omega \rho_\omega(x)G_\omega^0(x-x')\rho_\omega(x')}.
\end{eqnarray}
Using this recent relation and Eq.(\ref{2 point Green function})
we find the following expansion for Green's function
\begin{eqnarray}\label{cov.expansion Green function0}
G(x,x')&=&G^0(x-x') \nonumber \\
&+&\int_\Omega \int_\Omega  d^{n+1}z_1 d^{n+1}z_2  G^0(x-z_1)[\partial_{\mu_1} \partial_{\mu_2} \chi^{{\mu_1}{\mu_2}}
(z_1,z_2)]G^0(z_2-x') \nonumber\\
&+&\int_\Omega \int_\Omega \int_\Omega \int_\Omega d^{n+1}z_1 d^{n+1}z_2 d^{n+1}z_3 d^{n+1}z_4 G^0(x-z_1)
[\partial_{\mu_1} \partial_{\mu_2} \chi^{{\mu_1}{\mu_2}}(z_1,z_2)]G^0(z_2-z_3) \nonumber\\
&&G^0(z_3-z_4)[\partial_{\mu_3} \partial_{\mu_4}
\chi^{{\mu_3}{\mu_4}}(z_3,z_4)]G^0(z_4-x') +\cdots,
\end{eqnarray}
where susceptibility of the medium is given by
Eq.(\ref{relativistic susceptibility}). For a moving homogeneous
material, that is a medium for which the tensor
$\chi^{\mu_1,\mu_2}(x,x')$ is dependent only on the difference
$x-x'$, and the coupling function $f^\mu (\omega,x)$ is
independent of $x$, the Green's function can be written in a more
compact form in Fourier space as follows
\begin{equation}\label{cov Green}
G(k)=G_0(k)[\texttt{I}-G_0(k)k_{\mu_1}k_{\mu_2}\chi^{\mu_1 \mu_2}(k)].
\end{equation}
\section{Partition function }\label{Partition function}
\noindent The partition function of a real scalar field in
the presence of a medium is 
\begin{equation}\label{partition functions}
\Xi=\int D\varphi e^{\frac{i}{\hbar}S}=\int D\varphi e^{\frac{i}{\hbar} \int d^{n+1}x  {\cal L}},
\end{equation}
where ${\cal{L}}$ is given by Eq.(\ref{system lagrangian}). The partition function can also be written in frequency 
\cite{kapusta}
\begin{equation}\label{partition functions 1}
\Xi=\int D\varphi e^{-\frac{i}{2\hbar} \int \frac{d\omega}{2\pi} \int d^n\textbf{x}
\tilde{\varphi}(\textbf{x},-\omega)\left[-\omega^2 \frac{\epsilon(\omega,\textbf{x})}
{\epsilon_0}-\nabla^2 \right]\tilde{\varphi}(\textbf{x},\omega)},
\end{equation}
which is much suitable for our purposes. If we make a Wick's rotation $\omega=i\nu$ in frequency space, the action will be Euclidean and the free energy is given by $ E=-\frac{\hbar}{\tau} \ln \Xi$, where $\tau$ is the duration of interaction and usually is taken to be quite large. Now from standard path-integral techniques we find the free energy at finite temperature $T$ as
\begin{equation}\label{free energy}
E=\textit{k}_B T \sum _{l=0} ^\infty \ln \det[\hat{K}(\nu_l;\textbf{x},\textbf{x}')],
\end{equation}
where $\nu_l=2 \pi l \textit{k}_B T/\hbar$ are Matsubara
frequencies and $\textit{k}_B$ is Boltzman constant. Also
$\hat{K}(\nu_l;\textbf{x},\textbf{x}')=[\nu_l
\epsilon(i\nu_l,\textbf{x})-\nabla^2]\delta(\textbf{x}-\textbf{x}')$
and $\hat{K}(\nu_l;\textbf{x},\textbf{x}')=G^{-1}(i
\nu_l;\textbf{x},\textbf{x}')$. Therefore
\begin{equation}\label{free energy 2}
E=-\textit{k}_B T \sum _{l=0} ^\infty tr \ln[G(i\nu_l;\textbf{x},\textbf{x}')].
\end{equation}
Now using the expansion Eq.(\ref{expansion Green function}), we
obtain the expansion for free energy in terms of the
susceptibility function as follows
\begin{eqnarray}\label{expansion free energy}
E&=&\textit{k}_B T \sum_{l=0}^\infty \sum_{n=1}^\infty
\frac{(-1)^{n+1}}{n}\int d^n\textbf{x}_1\cdots d^n\textbf{x}_n G^0
(i\nu_l;\textbf{x}_1-\textbf{x}_2)...G^0(i\nu_l;\textbf{x}_n-\textbf{x}_1) \nonumber \\
&\times& \chi(i\nu_l,\textbf{x}_1)...\chi(i\nu_l,\textbf{x}_n),
\end{eqnarray}
where $G_0(i\nu,\textbf{x}-\textbf{x}')$ depends on the dimension
of the system.
\section{Examples}\label{Example}
\subsection{1+1 Dimension}\label{1+1 Dimension}
\noindent Now let us apply the previous formalism to obtain the force induced from a
fluctuating massless scalar field between two objects with
susceptibilities $\chi_1(\omega)$ and $\chi_2(\omega)$ in one dimension. In this case the fluctuating field is defined over a 
$(1+1)$-dimensional space-time $(x=(\textbf{x},t)\in\mathbb{R}^{1+1})$. In non-relativistic regime, the Green's function of the system and
reservoir are respectively given by
\begin{equation}\label{Green function 3}
G^0(\omega;\textbf{x}-\textbf{x}')=\frac{e^{-i\omega|\textbf{x}-\textbf{x}'|}}{2\omega},
\end{equation}
\begin{equation}\label{medium green}
G^0_{\omega'}(x-x')=\delta(\textbf{x}-\textbf{x}')\Theta(t-t')\frac{e^{-i{\omega}'(t-t')}}{2{\omega}'},
\end{equation}
and in relativistic regime, these Green's functions are respectively
 \begin{equation}\label{cov.Green}
G^0(x,x')=-\frac{i\hbar}{2\pi}\ln|x-x'|,
\end{equation}
\begin{equation}\label{cov.Green1}
G^0_\omega(x-x')=\frac{i\hbar}{2\pi}K_0(\omega |x-x'|),
\end{equation}
where $K_0(x)=\int _0 ^\infty dy \frac{\cos(xy)}{\sqrt{y^2 +1}},\hspace{0.5cm} (x>0)$, is the modified Bessel function.
Here we restrict ourselves to the non-relativistic regime and assume the following form for the dielectric function
\begin{eqnarray}\label{sucsyptibility}
\frac{\epsilon(\omega,\textbf{x})}{\epsilon_0}=\left\{
                                        \begin{array}{ll}
                                          \frac{\epsilon_1(\omega)}{\epsilon_0}, & \hbox{$a<\textbf{x}<b$} \\
                                          1, & \hbox{$b<\textbf{x}<c$} \\
                                          \frac{\epsilon_2(\omega)}{\epsilon_0}, & \hbox{$c<\textbf{x}<d.$}
                                        \end{array}
                                      \right.
\end{eqnarray}
By inserting Eq.(\ref{Green function 3}) into Eq.(\ref{expansion free energy}) and considering Eq. (\ref{sucsyptibility}),
up to the first relevant approximation, the free energy is obtained as
\begin{equation}\label{example E}
E=-K_B T \sum_{l=1} ^{\infty}\int d\textbf{x} \int d\textbf{x}' \frac{e^{-2\nu_l |\textbf{x}-\textbf{x}'|}}{(2\nu_l)^2}
\chi(i\nu_l,\textbf{x}) \chi(i\nu_2,\textbf{x}'),
\end{equation}
now for simplicity we assume that the susceptibilities are position independent, i.e., matter has been distributed in
$\Omega_1$ and $\Omega_2$ intervals homogeneously, then
\begin{equation}\label{example E1}
E=-K_B T\sum_{l=1}^{\infty} \chi(i \nu_l)\chi(i \nu_l)\int_{\Omega_1}d\textbf{x}
\int_{\Omega_2}d\textbf{x}'\frac{e^{-2\nu_l |\textbf{x}-\textbf{x}'|}}{(2\nu_l)^2},
\end{equation}
for further simplicity we assume that the objects have small dimensions compared to the distance between their centers of
masses, therefore
\begin{equation}\label{E f}
E=-k_B T \frac{(b-a)(d-c)}{4} \sum_{l=1}^{\infty} (\epsilon_1(i{\nu_l})-1)(\epsilon_2(i{\nu_l})-1) \frac{e^{-\nu_l(d+c-a-b)}}
{(2\nu_l)^2}.
\end{equation}
When the dimensions of objects are not necessarily smaller than the distance between their centers of masses we will find
the following expression for the free energy
\begin{eqnarray}\label{E}
E&=&-K_B T\sum_l[\frac{\chi^2_1(\nu_l)}{2\nu_l}(-2(b-a)+\frac{1}{\nu_l}-\frac{e^{-2\nu_l(b-a)}}{\nu_l})  \nonumber\\
&+&\frac{\chi^2_2(\nu_l)}{2\nu_l}(-2(d-c)+\frac{1}{\nu_l}-\frac{e^{-2\nu_l(d-c)}}{\nu_l})
-\frac{\chi_1(\nu_l) \chi_2(\nu_l)}{(\nu_l)^2}e^{-2\nu_l r}\sinh(2\nu_l r')\sinh(2\nu_l r'')],
\end{eqnarray}
where self energies are also included. By defining
\begin{eqnarray}\label{interval}
&a+b&=2r_1,\,\,\,\,\,\,\,\,\,b-a=2r'' \nonumber\\
&c+d&=2r_2,\,\,\,\,\,\,\,\,\,d-c=2r' ,\,\,\,\,\,\,r=r_2-r_1,
\end{eqnarray}
the force induced by the fluctuating field can be obtained as
\begin{equation}\label{force}
F=\frac{\partial E}{\partial r} = -K_B T \sum_l \frac{2\chi_1(\nu_l) \chi_2(\nu_l)}{\nu_l}e^{-2\nu_l r}\sinh(2\nu_l r')
\sinh(2\nu_l r'').
\end{equation}
In zero temperature and in the case that susceptibilities are independent of frequency, the force is given by
\begin{equation}
F=-\frac{\hbar \chi_1 \chi_2}{2\pi}(\Gamma[0,d-a]+\Gamma[0,c-b]-\Gamma[0,d-b]-\Gamma[0,c-a]),
\end{equation}
where $\Gamma[a,z]$ is the incomplete gamma function.
\subsection{2+1-Dimensional space-time}\label{2+1-Dimension}
\noindent In this section we restrict ourselves to a $(2+1)$-dimensional space-time $(x=(\textbf{x},t)\in\mathbb{R}^{2+1})$.
In this case and in the non-relativistic regime, the Green's function of the system is given by
\begin{equation}\label{2+1 Green}
G_0(\omega,\textbf{x}-\textbf{x}')=\frac{i \hbar}{2\pi}K_0(i\omega|\textbf{x}-\textbf{x}'|),
\end{equation}
and the form of the Green's function of the medium is the same as Eq.(\ref{medium green}) with the difference that now $\textbf{x}$ belongs to a two dimensional space. Now consider two objects with susceptibilities $\chi_1(\omega,\textbf{x})$ and $\chi_2(\omega,\textbf{x})$, the free energy in
first approximation is
\begin{equation}\label{2+1 Free energy}
E=-\frac{K_B T}{4{\pi}^2} \sum_l ^\infty \int d^2\textbf{x} \int d^2\textbf{x}' K_0^2(\nu_l|\textbf{x}-\textbf{x}'|)
\chi_1(i\nu_l,\textbf{x})\chi_2(i\nu,\textbf{x}'),
\end{equation}
where self energies are ignored. In zero temperature the summation over the positive integer $l$ is replace by an integral
according to the rule $\hbar \int _0 ^\infty \frac{d \nu}{2 \pi}\rightarrow K_B T \sum _{l=0}^\infty$, therefore, the free
energy in zero temperature is
\begin{equation}\label{2+1 T=0}
E=-\frac{1}{32{\pi}^4}\int d\nu \int d^2\textbf{x} \int d^2\textbf{x}' K_0^2(\nu|\textbf{x}-\textbf{x}'|)
\chi_1(i\nu_l,\textbf{x})\chi_2(i\nu,\textbf{x}').
\end{equation}
In the case that the susceptibilities are independent of the frequency $\nu$, we obtain
\begin{equation}\label{example 2+1}
E=-\frac{1}{32{\pi}^3} \int d^2 \textbf{x} \int d^2 \textbf{x}' \frac{\chi_1(\textbf{x})\chi_2(\textbf{x})}
{|\textbf{x}-\textbf{x}'|^2}.
\end{equation}
Using this recent formula we can obtain for example the Casimir energy of two rings \cite{milton}. Assume that the two ring
have radii $a$ and $b$, and their centers are separated by a distance $R$, Fig.(1). This situation can be shown by
susceptibilities $\chi_1(\textbf{x})=\chi_1 \delta(r-a)$ and $\chi_2(\textbf{x})=\chi_2 \delta(r'-b)$ where $r$ and $r'$
are radial coordinates in cylindrical polar coordinate system. By using Eq.(\ref{example 2+1}) the free energy for this
geometry is given by
\begin{equation}\label{E 2+1}
E=\frac{-\chi_1 \chi_2 a b}{32{\pi}^3}\int_0^{2\pi}d\theta \int_0^{2\pi}d\theta' \frac{1}{R^2+a^2+b^2-2a R \cos\theta +
2bR \cos\theta'-2ab \cos(\theta-\theta')}.
\end{equation}
With a simple change in angular coordinates to $u=\theta-\theta'$ and $v=\frac{\theta+\theta'}{2}$ this expression can be
integrated to yield the exact closed form \cite{wagner}
\begin{equation}\label{E.F 2+1}
E=-\frac{\chi_1 \chi_2 a b}{8\pi}\frac{1}{\sqrt{(R^2-(a-b)^2)(R^2-(a+b)^2)}},
\end{equation}
which is the same result obtained from the with weak-coupling regime in scattering formalism \cite{milton, kimball}.
In relativistic regime, the Green's functions of the system and reservoir respectively are
\begin{eqnarray}\label{1+2 cov green}
G_0(x-x')=\frac{1}{4\pi}\frac{1}{|x-x'|}, \nonumber\\
G_0^\omega(x-x')=\frac{1}{4\pi}\frac{e^{-i\omega|x-x'|}}{|x-x'|},
\end{eqnarray}
and a similar approach can be followed for obtaining relativistic results.
\subsection{3+1-Dimensional space-time}\label{3+1-Dimansion}
\noindent In this section we restrict ourselves to a $(3+1)$-dimensional space-time $(x=(\textbf{x},t)\in\mathbb{R}^{3+1})$.
In this case, in non-covariant regime, the Green's function of the system is given by
\begin{equation}\label{1+3 dimension}
G_0(\omega,\textbf{x}-\textbf{x}')=\frac{1}{4\pi}\frac{e^{-i\omega|\textbf{x}-\textbf{x}'|}}{|\textbf{x}-\textbf{x}'|},
\end{equation}
and the free energy in zero temperature and in the first order of approximation is
\begin{equation}\label{3+1 E.F}
E=-\frac{1}{64{\pi}^3}\int _{-\infty} ^\infty d\nu \int d^3\textbf{x} \int d^3 \textbf{x}' \frac{e^{-2\nu|\textbf{x}-
\textbf{x}'|}\chi_1(\textbf{x},\nu)\chi_2(\textbf{x}',\nu)}{|\textbf{x}-\textbf{x}'|^2},
\end{equation}
where again we have ignored the self energies which are irrelevant here. If the susceptibilities are independent of frequency
the expression simplifies further to
\begin{equation}\label{3+1 E.1}
E=-\frac{1}{64{\pi}^3}\int d^3\textbf{x} \int d^3\textbf{x}' \frac{\chi_1(\textbf{x})\chi_2(\textbf{x}')}{|\textbf{x}-
\textbf{x}'|^3}.
\end{equation}
At finite temperature the integral over frequency becomes the Matsubara sum, so the energy becomes
\begin{equation}\label{3+1 E}
E=-\frac{K_B T}{32{\pi}^2}\int d^3\textbf{x}\int d^3\textbf{x}' \chi_1(\textbf{x})\chi_2(\textbf{x}')\frac{\coth 2\pi
T|\textbf{x}-\textbf{x}'|}{|\textbf{x}-\textbf{x}'|}.
\end{equation}
As an example, we consider two spheres, of radius $a$ and $b$, respectively, with a distance between their centers $R>a+b$.
The susceptibilities of the spheres are $\chi_1(\textbf{x})=\chi_1 \delta(r-a)$ and $\chi_2(\textbf{x})=\chi_2 \delta(r'-b)$,
where $r$ and $r'$ are radial coordinates in spherical coordinate systems and $R$ lies along the $z$ axis of both coordinate
systems, Fig.(2). Then the distance between points on the spheres is
\begin{eqnarray}\label{distance}
|\textbf{x}-\textbf{x}'|=\sqrt{R^2+a^2+b^2-2ab \cos\gamma-2 R(a \cos\theta-b \cos\theta')}, \nonumber\\
 \cos\gamma=\cos\theta \cos\theta'+\sin\theta \sin\theta' \cos(\phi-\phi').
\end{eqnarray}
By inserting Eq.(\ref{distance}) into Eq.(\ref{3+1 E.1}) energy in zero temperature is given by
\begin{equation}\label{exapmle 3+1}
E=-\frac{\chi_1 \chi_2 a b}{16 \pi R}\ln {\frac{1-(a-b)^2/R^2}{1+(a+b)^2/R^2}},
\end{equation}
which is the same result obtained from the scattering method \cite{kimball}.
\section{ELECTROMAGNETIC FIELD}\label{ELECTROMAGNETIC FIELD}
\noindent Now we find the Green's function of the electromagnetic field in the presence of some dielectrics from which the Casimir force in quite complicated geometries can be obtained approximately. For this purpose, the total Lagrangian density can be written in Coulomb gauge $\nabla\cdot\textbf{A}=0, \, A^0=0$, as follows
\cite{soltani}
\begin{equation}\label{1}
{\cal{L}}=\frac{1}{2}(\textbf{E}^2-\textbf{B}^2)+\frac{1}{2}\int_0^\infty d\omega (\dot{\textbf{Y}}_\omega ^2(x)-\omega^2 \textbf{Y}_\omega ^2(x))+\int d\omega f(\omega,\textbf{x})\textbf{A}.\dot{\textbf{Y}}_\omega.
\end{equation}
Now the interacting generating functional is given by
\begin{eqnarray}\label{1.W,EM}
W&=&\int D[\textbf{A}]\prod_\omega D[\textbf{Y}_\omega]\exp\frac{i}{\hbar}\int d^4x [-\frac{1}{2}A_\mu
\hat{K}_{ij}A_j-\int_0^\infty d\omega \frac{1}{2} Y_{\omega,i}(\partial_t^2+\omega^2)\delta_{ij}Y_{\omega,j} \nonumber\\
&+&\int _0^\infty d\omega f(\omega,\textbf{x})A_i\dot{Y}_{\omega,i}+ J_i A_i +\int_0^\infty d\omega J_{\omega,i}
Y_{\omega,i}],
\end{eqnarray}
where the kernel $\hat{K}_{ij}$ is defined by
\begin{equation}\label{kernel}
\hat{K}_{ij}=(\frac{\partial_0^2}{c^2}-\nabla^2)\delta_{i,j}-\partial_i \partial_j.
\end{equation}
Now from the well known relation
\begin{equation}\label{G}
G_{ij}(x,x')=(\frac{\hbar}{i})^2\frac{\delta^2}{\delta J_i(x)\delta J_j(x')} W[j,{j_\omega}]|_{j,{j_\omega}=0},
\end{equation}
and following the same process we did for the scaler case, we obtain the following expansion for Green's function in
frequency 
\begin{eqnarray}\label{expansion Green function,EM}
G_{ij}(\textbf{x}-\textbf{x}',\omega)=G^0_{ij}(\textbf{x}-\textbf{x}',\omega)+ \int_\Omega d^3 \textbf{z}_1 G^0_{il}
(\textbf{x}-\textbf{z}_1,\omega)[\omega^2 \tilde{\chi}(\omega,\textbf{z}_1)]G^0_{lj}(\textbf{z}_1-\textbf{x}',\omega)+
\nonumber\\
\int_\Omega \int_\Omega d^3 \textbf{z}_1 d^3 \textbf{z}_2 G^0_{il}(\textbf{x}-\textbf{z}_1,\omega)[\omega^2
\tilde{\chi}(\omega,\textbf{z}_1)]G^0_{lm}(\textbf{z}_1-\textbf{z}_2,\omega)[\omega^2
\tilde{\chi}(\omega,\textbf{z}_2)]G^0_{mj}(\textbf{z}_2-\textbf{x}',\omega) +\cdots,\nonumber\\
\end{eqnarray}
It can be easily shown that Green's function (\ref{expansion Green function,EM}) satisfies the follows
equation \cite{kheirandish1}
\begin{equation}\label{EM.EQN}
[-\frac{\omega^2}{c^2}\epsilon(\omega,\textbf{x})\delta_{ij}-\nabla^2\delta_{ij}+\partial_i\partial_j]G_{ij}
(\textbf{x},\textbf{x}',\omega)=
\delta^3(\textbf{x}-\textbf{x}')\delta_{ij},
\end{equation}
where $\frac{\epsilon(\omega)}{\epsilon_0}=\chi(\omega)+1$.
A similar approach can be followed to find the partition function in terms of the susceptibility of the medium as follows
\begin{eqnarray}\label{expansion free energy.EM}
E &=& \textit{k}_B T \sum_{l=0}^\infty \sum_{n=1}^\infty \frac{(-1)^{n+1}}{n}\int d^3\textbf{x}_1\cdots d^3
\textbf{x}_n G^0_{i_1i_2}(i\nu_l;\textbf{x}_1-\textbf{x}_2)...G^0_{i_ni_1}(i\nu_l;\textbf{x}_n-\textbf{x}_1) \nonumber \\
& \times & \chi(i\nu_l,\textbf{x}_1)\cdots\chi(i\nu_l,\textbf{x}_n),
\end{eqnarray}
where the free Green's function $G^0_{ij}(\textbf{x}-\textbf{x}',i\nu_l)$ satisfies Eq.(\ref{EM.EQN}) with
$\epsilon(\omega,\textbf{x})=1$ and $\omega=i\nu_l$. By defining $\textbf{r}=\textbf{x}-\textbf{x}'$, we find
\begin{equation}\label{G.EM}
G_{ij}^0(\textbf{r},i\nu_l)=\frac{\nu_l^2}{c^2}\frac{e^{-\frac{\nu_lr}{c}}}{4\pi r}[\delta_{ij}(1+\frac{c}{\nu_lr}+
\frac{c^2}{\nu_l^2 r^2})-\frac{r_ir_j}{r^2}(1+\frac{3c}{\nu_l r}+\frac{3c^2}{\nu_l^2 r^2})]+\frac{1}{3}\delta_{ij}
\delta^3(\textbf{r}).
\end{equation}
As an example of the application of Eq.(\ref{expansion free energy.EM}), let us find the interaction energy of a system
composed of two dielectrics with volumes $V_1$ and $V_2$ and the susceptibilities $\chi_1$ and $\chi_2$, respectively.
The first relevant nonzero term corresponds to $n=2$, therefore
\begin{equation}\label{E example}
E=-\frac{1}{2}k_B T\sum_{l=0}^\infty \int_{V_1}\int_{V_2} d^3\textbf{x} d^3\textbf{x}' G^0_{ij}(\textbf{x}-
\textbf{x}',i\nu_l)G^0_{ji}(\textbf{x}'-\textbf{x},i\nu_l)
\chi_1(i\nu_l,\textbf{x})\chi_2(i\nu_l,\textbf{x}').
\end{equation}
Inserting the Green's function (\ref{G.EM}) into (\ref{E example}), we find
\begin{equation}\label{E.EM}
E=-k_B T\sum_{l=0}^\infty \int_{V_1}\int_{V_2} d^3\textbf{x} d^3\textbf{x}' \chi_1(i\nu_l,\textbf{x})
\chi_2(i\nu_l,\textbf{x}')h(\nu_l,|\textbf{x}-\textbf{x}'|),
\end{equation}
where we have defined
\begin{equation}\label{h}
h(\nu_l,|\textbf{x}-\textbf{x}'|)=\frac{e^{-\frac{2\nu_l}{c}|\textbf{x}-\textbf{x}'|}}{8{\pi}^2}
\{\frac{(\frac{\nu_l}{c})^4}{|\textbf{x}-\textbf{x}'|^2}
+\frac{2(\frac{\nu_l}{c})^3}{|\textbf{x}-\textbf{x}'|^3}+\frac{5(\frac{\nu_l}{c})^2}{|\textbf{x}-\textbf{x}'|^4}
+\frac{6\frac{\nu_l}{c}}{|\textbf{x}-\textbf{x}'|^5}+\frac{3}{|\textbf{x}-\textbf{x}'|^6} \}.
\end{equation}
In zero temperature, the summation over the positive integer $l$ is replaced by an integral, therefore for the case
that susceptibilities are independent of frequency, the free energy is given by
\begin{equation}\label{FE}
E=-\frac{23}{64 {\pi}^3}\int d^3\textbf{x} \int d^3\textbf{x}' \frac{\chi_1(\textbf{x}) \chi_2(\textbf{x}')}
{|\textbf{x}-\textbf{x}'|^7}.
\end{equation}
For example for two spheres of radii $a$ and $b$, the distance
between their centers $R>a+b$, with susceptibilities
$\chi_1(\textsl{x})=\chi_1 \delta(r-a)$ and
$\chi_2(\textbf{x})=\chi_2 \delta(r'-b)$, where $r$ and $r'$ are
radial coordinates in spherical coordinate systems, the free
energy from Eq.(\ref{FE}) is as follows
\begin{equation}\label{FE.example}
E=-\frac{23 \chi_1 \chi_2 a^2 b^2}{64 {\pi}^3}\int d\Omega \int d\Omega' \frac{1}{|\textbf{x}-\textbf{x}'|^7}.
\end{equation}
Now we use the following formula [?]
\begin{equation}\label{integral}
\int d\Omega d\Omega' |\textbf{x}-\textbf{x}'|^p=(4\pi)^2 R^p P_p(\hat{a},\hat{b}),
\end{equation}
where $R$ is the distance between the centers of the two non-overlapping spheres of radii $a$ and $b$, respectively.
Here $\hat{a}=a/R$ and $\hat{b}=b/R$, and $P_p(\hat{a},\hat{b})$ can in general be represented by the infinite series
\begin{eqnarray}\label{integral c}
P_p(\hat{a},\hat{b})=\sum_{n=0}^\infty \frac{2}{(2n+2)!}\frac{\Gamma(2n-p-1)}{\Gamma(-p-1)}Q_n(\hat{a},\hat{b}),\nonumber\\
Q_n=\frac{1}{2}\sum_{m=0}^n {\left( {\begin{array}{*{20}c}
   {2n + 2}  \\
   {2m + 1}  \\
\end{array}} \right)}\hat{a}^{2(n-m)}\hat{b}^{2m}.
\end{eqnarray}
We can easily see the following recursion relation holds:
\begin{equation}\label{recursion}
 P_{p-1}(\hat{a},\hat{b})=\frac{R^{-p}}{1+p}\frac{\partial}{\partial R}R^{p+1}P_p(\hat{a},\hat{b}),\,\,\,\,P_{-1}=1.
\end{equation}
Using this recent relation, the free energy in zero temperature can be written as
\begin{equation}\label{FE.example1}
E_0=-\frac{23 \chi_1 \chi_2 a^2 b^2}{4\pi R^7}P_{-7},
\end{equation}
where
\begin{eqnarray}\label{11}
P_{-7}&=&\frac{-2 (\hat{a}^2 - \hat{b}^2)^4 ( \hat{a}^2 + \hat{b}^2-5) + (\hat{a}^2 - \hat{b}^2)^2 (
    52 (\hat{a}^2 + \hat{b}^2-44) - 24 (\hat{a}^2 + \hat{b}^2)^2)}{10 (\hat{a}^4 + (\hat{b}^2-1)^2 - 2
    \hat{a}^2 (\hat{b}^2+1))^4}  \nonumber\\
&+&\frac{ 2 (5 - 5 (\hat{a}^2 + \hat{b}^2) + 8 (\hat{a}^2 + \hat{b}^2)^2 -
    4 (\hat{a}^2 + \hat{b}^2)^3)}{10 (\hat{a}^4 + (\hat{b}^2-1)^2 - 2 \hat{a}^2 (\hat{b}^2+1))^4}.
\end{eqnarray}
In non zero temperature and first order of correction, the free energy is
\begin{equation}\label{E_t}
 E=-\frac{23 \chi_1 \chi_2 a^2 b^2}{4\pi R^7}P_{-7}(\hat{a},\hat{b})-\frac{6 k_B T \chi_1 \chi_2 a^2 b^2}{R^6}
 P_{-6}(\hat{a},\hat{b}),
\end{equation}
where
\begin{equation}\label{p-6}
P_{-6}(\hat{a},\hat{b})=\frac{1}{4\hat{a}\hat{b}}[\frac{1}{(\hat{a} - \hat{b} -1)^3} -\frac{1}{(\hat{a} +
\hat{b}-1)^3} -\frac{1}{(\hat{a}-\hat{b}+ 1)^3} + \frac{1}{(\hat{a} + \hat{b} +1)^3}].
\end{equation}
It can be shown that this results are equivalent with the finite temperature weak coupling regime in scattering
formalism \cite{wagner,kimball}.
\subsection{Covariant formulation}\label{Covariant formulation}
\noindent For a covariant formulation we consider the following
Lorentz invariant Lagrangian density in Lorentz gauge
\begin{equation}\label{cov.lagrangian}
\textit{L}(x)=-\frac{1}{4}F_{\mu,\nu}F^{\mu,\nu}+ \frac{1}{2}\int _0^\infty d\omega
[\partial_\mu Y_\omega \partial^\mu Y_\omega-\omega^2 Y_\omega^2]+\int_0^\infty d\omega f^{\mu \nu}
(\omega,x)Y_\omega \partial_\mu A_\nu,
\end{equation}
where $F^{\mu,\nu}=\partial_\mu A_\nu -\partial^\nu A^\mu$ and
$f^{\mu,\nu}(\omega,\textsl{x})$ is an antisymmetric coupling
tensor which couples electromagnetic field to the medium. From
Euler-Lagrange equations we can show that Green's function
satisfies the following equation \cite{kheirandish1}
\begin{equation}\label{cov.green1}
\Box G_{\mu \nu}(x-x')-\int d^4 x'' g_{\mu \delta}\partial_\gamma \partial_\alpha^{''}
\chi^{\gamma\delta\alpha\beta}(x,x'')G_{\beta\nu}(x''-x)=g_{\mu\nu}\delta^4(x-x'),
\end{equation}
where the susceptibility tensor $\chi^{\nu\mu\alpha\beta}(x,x')$ is defined by
\begin{equation}\label{sus}
\chi_{\nu\mu\alpha\beta}(x,x')=\int_0^\infty d\omega
f^{\mu\nu}(\omega,x)G^0_\omega(x-x')f^{\alpha\beta}(\omega,x').
\end{equation}
For Green's function we find the following expansion in term of
the susceptibility tensor
\begin{eqnarray}\label{cov.g}
G_{\mu \nu}(x,x')=G_{\mu,\nu}^0(x-x')+\int d^4 z_1 d^4 z_2 G_{\mu \nu_1}^0(x-z_1)\Gamma^{\nu_1 \nu_2}
(z_1,z_2)G_{\nu_2 \nu}^0(z_2-x') \nonumber\\
\int d^4 z_1... d^4 z_4 G_{\mu \nu_1}^0(x-z_1)\Gamma^{\nu_1
\nu_2}(z_1,z_2)G_{\nu_2 \nu_3}^0(z_2-z_3) \Gamma^{\nu_3
\nu_4}(z_3,z_4)G_{\nu_4 \nu}^0(z_4-x')+\cdots,
\end{eqnarray}
where we have defined $\Gamma ^{\nu_1
\nu_2}(z_1,z_2)=\partial_{\mu_1}\partial_{\mu_2} \chi^{\mu_1 \nu_1
\mu_2 \nu_2}(z_1,z_2)$. If we are given the susceptibility tensor
of the medium, we can find the Green's function perturbatively in
terms of the susceptibility. Having Green's function, we can
investigate for example the dynamical energy configurations, which
is closely related to the dynamical Casimir effects.
\section{Proca Electromagnetic Field}\label{proca Electromagnetic Field}
\noindent The Proca equation is a relativistic wave equation for a
massive spin-1 particle. The weak interaction is transmitted by
such kind of vector bosons \cite{greiner1,itzykson}. The
transition from the massless electromagnetic field to a field of
massive vector bosons is easily done. In contrast to the
electromagnetic field, the massive spin-1 field will automatically
satisfy the Lorentz condition $\partial_\nu A^\nu=0$ and the
Lorentz condition has become a condition of consistency for the
Proca field. Therefore the Lagrangian density of the system is
given by
\begin{equation}\label{p.L}
{\cal{L}}=-\frac{1}{2}F_{\mu \nu}F^{\mu \nu}+\frac{m^2c^2}{2\hbar} A_\mu A^\mu+
\frac{1}{2}\int_0^\infty d\omega (\dot{\textbf{Y}}_\omega ^2(x)-\omega^2
\textbf{Y}_\omega ^2(x))+\int d\omega f(\omega,\textbf{x})\textbf{A}.\dot{\textbf{Y}}_\omega.
\end{equation}
The interacting generating functional can be written in terms of the vector potential and the medium fields as
\begin{eqnarray}\label{P.W,EM}
W&=&\int D[\textbf{A}]\prod_\omega D[\textbf{Y}_\omega]\exp\frac{i}{\hbar}\int
d^4x [-\frac{1}{2}A_\mu \hat{K}_{\mu \nu}A_\nu-\int_0^\infty d\omega \frac{1}{2}
Y_{\omega,\mu}(\partial_t^2+\omega^2)\delta_{\mu \nu}Y_{\omega,\nu} \nonumber\\
&+&\int _0^\infty d\omega f(\omega,\textbf{x})A_\mu\dot{Y}_{\omega,\mu}+
J_\mu A_\mu +\int_0^\infty d\omega J_{\omega,\mu}Y_{\omega,\mu}],
\end{eqnarray}
where the kernel $\hat{K}_{\mu \nu}$ is defined by
\begin{equation}\label{P.kernel}
\hat{K}_{\nu \mu}=[\frac{\partial_0^2}{c^2}-\nabla^2]\delta_{\mu \nu}-\partial_\mu \partial_\nu-
\frac{m^2 c^2}{\hbar^2} \delta_{\mu \nu}.
\end{equation}
Now from the well known relation
\begin{equation}\label{P.G}
G_{\nu \mu}(x,x')=(\frac{\hbar}{i})^2\frac{\delta^2}{\delta J_\nu(x)\delta J_\mu(x')} W[j,{j_\omega}]|_{j,{j_\omega}=0},
\end{equation}
and following the same process we did for the scaler field, we
obtain the following expansion for Green's function in frequency
variable
\begin{eqnarray}\label{P.expansion Green function,EM}
G_{\mu \nu}(\textbf{x}-\textbf{x}',\omega)=G^0_{\mu \nu}(\textbf{x}-\textbf{x}',\omega)+
\int_\Omega d^3 \textbf{z}_1 G^0_{\mu \alpha}(x-z_1,\omega)[\omega^2 \tilde{\chi}(\omega,\textbf{z}_1)]
G^0_{\alpha \nu}(\textbf{z}_1-\textbf{x}',\omega)+ \nonumber\\
\int_\Omega \int_\Omega d^3 \textbf{z}_1 d^3 \textbf{z}_2 G^0_{\mu
\alpha}(\textbf{x}-\textbf{z}_1,\omega) [\omega^2
\tilde{\chi}(\omega,\textbf{z}_1)]G^0_{\alpha\beta}(\textbf{z}_1-\textbf{z}_2,\omega)
[\omega^2 \tilde{\chi}(\omega,\textbf{z}_2)]G^0_{\beta
\nu}(\textbf{z}_2-\textbf{x}',\omega) +\cdots.\nonumber\\
\end{eqnarray}
Now using Euler-lagrange Equations, we find the equation of motion
for the vector potential as
\begin{equation}\label{P.eq}
\hat{K}_{\mu \nu}A_\nu+\frac{\partial}{\partial t}\int_0^\infty \frac{d\omega'}{2 \pi} \tilde{\chi}
(\omega',\textbf{x})\int dt' e^{i\omega'(t-t')}\frac{\partial}{\partial t'} A_\nu(t')=\int_0^\infty
d\omega f(\omega,\textbf{x})\dot{Y}_{\omega,\nu}^N(\omega,\textbf{x}),
\end{equation}
where $Y_{\omega,i}^N$ is as a noise or fluctuating field which
does not affect the Green's function. The Green's function of the
above equation satisfies
\begin{equation}\label{P.GF}
[-(\frac{\omega^2}{c^2}\epsilon(\omega,\textsl{x})+\frac{m^2c^2}{\hbar^2})\delta_{\mu \nu}-\nabla^2
\delta_{\mu \nu}-\partial_\mu \partial_\nu]G_{\mu \nu}(\textsl{x},\textsl{x}',\omega)=\delta^3(\textsl{x}-
\textsl{x}')\delta_{\mu \nu},
\end{equation}
where
$\epsilon(\omega,\textbf{x})=\epsilon_0[1+\tilde{\chi}(\omega,\textbf{x})]$.
It can be easily shown that Green's function (\ref{P.expansion
Green function,EM}) satisfies in Eq.(\ref{P.GF}). A similar
approach can be followed to find the free energy in terms of the
susceptibility of the medium as
\begin{eqnarray}\label{P.expansion free energy}
E&=&\textit{k}_B T \sum_{l=0}^\infty \sum_{n=1}^\infty \frac{(-1)^{n+1}}{n}\int d^3\textbf{x}_1 \cdots d^3
\textbf{x}_n G^0_{\nu_1\nu_2}(i\nu_l;\textbf{x}_1-\textbf{x}_2)\cdots G_{\nu_n \nu_1}^0(i\nu_l;\textbf{x}_n-
\textbf{x}_1) \nonumber \\
&\times& \chi(i\nu_l,\textbf{x}_1)\cdots
\chi(i\nu_l,\textbf{x}_n),
\end{eqnarray}
where the free Green's function $G_{\mu
\nu}^0(\textbf{x}-\textbf{x}',i\nu_l)$ satisfies in
Eq.(\ref{P.GF}) with $\epsilon(\omega,\textbf{x})=1$ and
$\omega=i\nu_l$. By defining $\textbf{r}=\textbf{x}-\textbf{x}'$,
we find
\begin{equation}\label{p.G0}
G_{\mu \nu}^0(\textbf{r},i\nu_l)=\zeta^2 \frac{e^{-\zeta r}}{4 \pi r}\left[ \delta_{\mu \nu}(1+\frac{1}{\zeta r}+
\frac{1}{\zeta^2 r^2})-\frac{r_\mu r_\nu}{r^2} (1+\frac{3}{\zeta r}+ \frac{3}{\zeta^2 r^2}) \right]+\frac{1}{3}
\delta_{\mu \nu} \delta^3(\textbf{r}),
\end{equation}
where $\zeta=\sqrt{(\frac{\nu_l}{c})^2-(\frac{mc}{\hbar})^2}$. As
an example of applications of Eq.(\ref{P.expansion free energy}),
let us find the interaction energy of a system composed of two
dielectrics with the volumes $V_1$ and $V_2$ and susceptibilities
$\chi_1$ and $\chi_2$, respectively. The first relevant nonzero
term corresponds to $n=2$, therefore
\begin{equation}\label{E example.p}
E=-\frac{1}{2}k_B T\sum_{l=0}^\infty \int_{V_1}\int_{V_2} d^3\textbf{x} d^3\textbf{x}'
G^0_{\mu \nu}(\textbf{x}-\textbf{x}',i\nu_l)G^0_{\nu \mu}(\textbf{x}'-\textbf{x},i\nu_l)
\chi_1(i\nu_l,\textbf{x})\chi_2(i\nu_l,\textbf{x}').
\end{equation}
Inserting the Green's function (\ref{p.G0}) into (\ref{E example.p}), we find
\begin{equation}\label{E.EM}
E=-k_B T\sum_{l=0}^\infty \int_{V_1}\int_{V_2} d^3\textbf{x} d^3\textbf{x}' \chi_1(i\nu_l,\textbf{x})
\chi_2(i\nu_l,\textbf{x}')h(\nu_l,|\textbf{x}-\textbf{x}'|),
\end{equation}
where we have defined
\begin{equation}\label{h}
h(\nu_l,|\textbf{x}-\textbf{x}'|)=\frac{e^{-2\zeta|\textbf{x}-\textbf{x}'|}}{8{\pi}^2}
\{\frac{\zeta^4}{|\textbf{x}-\textbf{x}'|^2}
+\frac{2 \zeta^3}{|\textbf{x}-\textbf{x}'|^3}+\frac{5 \zeta^2}{|\textbf{x}-\textbf{x}'|^4}
+\frac{6 \zeta}{|\textbf{x}-\textbf{x}'|^5}+\frac{3}{|\textbf{x}-\textbf{x}'|^6} \}.
\end{equation}
In zero temperature we have
\begin{eqnarray}\label{p.E}
E=-\frac{c}{32\pi^4}\int _{V_1}\int_{V_2} d^3\textbf{x} d^3 \textbf{x}'  \chi_1(\textbf{x})
\chi_2(\textbf{x})[(\frac{mc}{\hbar})^5 \frac{1}{|\textbf{x}-\textbf{x}'|^2}G^{31}_{13}
\left((\frac{mc}{\hbar})^2 |\textbf{x}-\textbf{x}'|^2\mid^{-2}_{\frac{-5}{2},0,\frac{1}{2}}\right)\nonumber\\
+ 2(\frac{mc}{\hbar})^4 \frac{1}{|\textbf{x}-\textbf{x}'|^3}G^{31}_{13}\left((\frac{mc}{\hbar})^2 |
\textbf{x}-\textbf{x}'|^2\mid^{-\frac{3}{2}}_{-2,0,\frac{1}{2}}\right) \nonumber\\
+ 5(\frac{mc}{\hbar})^3 \frac{1}{|\textbf{x}-\textbf{x}'|^4}G^{31}_{13}\left((\frac{mc}{\hbar})^2
|\textbf{x}-\textbf{x}'|^2\mid^{-1}_{-\frac{3}{2},0,\frac{1}{2}}\right) \nonumber\\
+ 6(\frac{mc}{\hbar})^2 \frac{1}{|\textbf{x}-\textbf{x}'|^5}G^{31}_{13}\left((\frac{mc}{\hbar})^2
|\textbf{x}-\textbf{x}'|^2\mid^{-\frac{1}{2}}_{-1,0,\frac{1}{2}}\right) \nonumber\\
+ 3(\frac{mc}{\hbar}) \frac{1}{|\textbf{x}-\textbf{x}'|^6}G^{31}_{13}\left((\frac{mc}{\hbar})^2 |
\textbf{x}-\textbf{x}'|^2\mid^{0}_{-\frac{1}{2},0,\frac{1}{2}}\right)]
\end{eqnarray}
where $G^{m,n}_{p,q}\left(x|^{a_1,...,a_n}_{b_1,...,b_q} \right)$
is the Meijer function \cite{Meijer}. Now for simplicity let us
assume that the volumes have small dimensions compared to the
distance $R$ between their centers of masses, such that $|r_1 -
r_2| \approx R$. Therefore
\begin{eqnarray}
E=-\frac{c\chi_1 \chi_2 V_1 V_2}{\pi^3} &[&\frac{23}{64 R^7}-\frac{mc}{\hbar}\frac{3}
{16 R^6}+(\frac{mc}{\hbar})^2 \frac{3}{64 R^5}(-3+2\gamma+2 \log R) \nonumber\\
&+&(\frac{mc}{\hbar})^3 \frac{1}{48
R^4}+(\frac{mc}{\hbar})^4\frac{6}{R^3}(-1+4\gamma+4\log
R)+\cdots],
\end{eqnarray}
where $\gamma=0.577$. The first term gives the normal Casimir
attractive force and the lowest order finite-mass correction tends
to reduce the effect.
\section{conclusion}\label{conclusion}
\noindent Quantum field theory in the presence of a medium is investigated for a scalar field and massless and massive vector
fields in different dimensions. Finite temperature corrections are considered and series expansions for corresponding Green's
functions and free energies are obtained in terms of the susceptibility function of the medium. Some illustrative examples
are given showing the applicability and the efficiency of the method. It is shown that the Casimir energy in first order
approximation in the present method is equivalent with the weak coupling regime in the scattering method. The covariant formulation of
the problem is presented. Finally, the Casimir energy in the fluctuating Proca field is investigated.
\bibliographystyle{apsrev4-1}

\newpage
\begin{figure}
\begin{center}
\includegraphics[width=2in, height=2.5in]{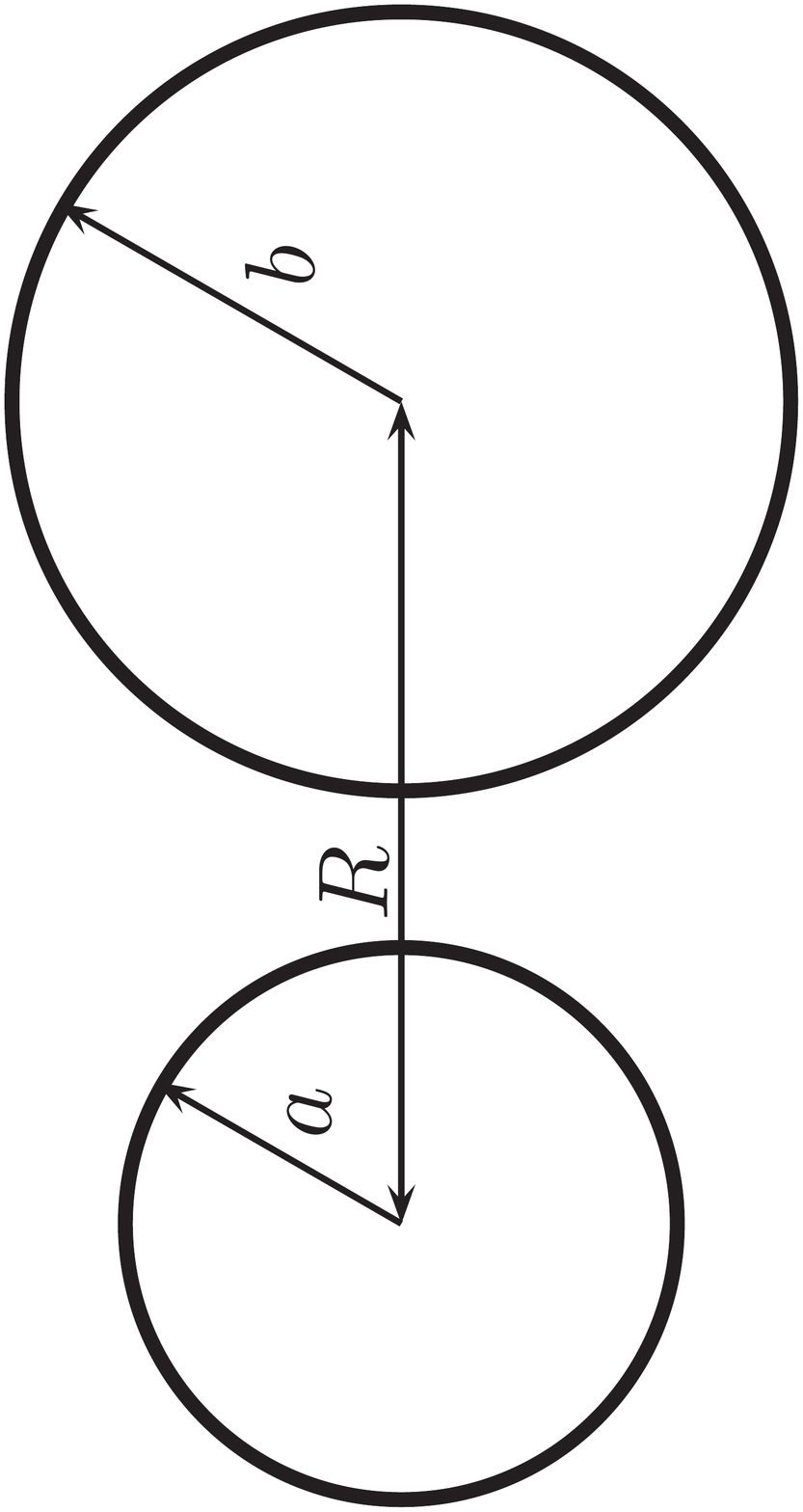}
\end{center}
\end{figure}
\textbf{Fig.1:} Two cylinders of radii $a$ and $b$, their centers are separated by a distance $R>a+b$.\\
\begin{figure}
\begin{center}
\includegraphics[width=3in, height=2.5in]{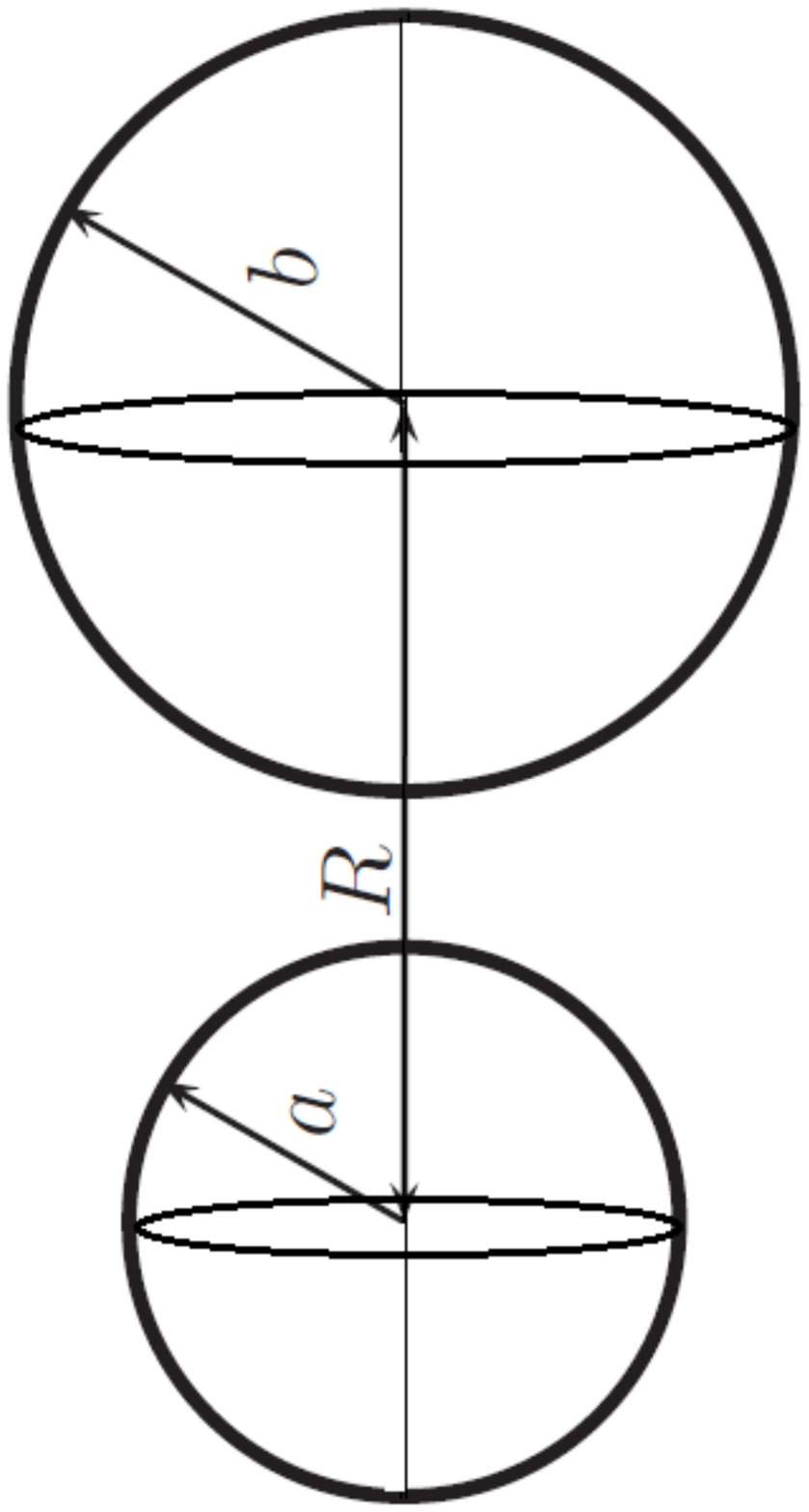}
\end{center}
\end{figure}
\textbf{Fig.2:} Two spheres of radii $a$ and $b$, their centers are separated by a distance $R>a+b$.\\
\end{document}